\pdfoutput=1
\documentclass[iop, apj, twocolappendix, numberedappendix]{emulateapj}

\usepackage{xspace}
\usepackage{amsmath}
\usepackage{framed} 
\usepackage{txfonts}
\usepackage{soul}
\usepackage{epstopdf}
\usepackage{xcolor}
\usepackage{rotating}
\usepackage{natbib}
\usepackage{ulem}
\usepackage{xspace}
\usepackage[colorlinks=true,urlcolor=black,linkcolor=blue,citecolor=blue]{hyperref}
\usepackage{sistyle}

\def\gtsima{$\; \buildrel > \over \sim \;$}
\def\ltsima{$\; \buildrel < \over \sim \;$}
\def\prosima{$\; \buildrel \propto \over \sim \;$}
\def\gsim{\lower.7ex\hbox{\gtsima}}
\def\lsim{\lower.7ex\hbox{\ltsima}}
\def\simgt{\lower.7ex\hbox{\gtsima}}
\def\simlt{\lower.7ex\hbox{\ltsima}}
\def\simpr{\lower.7ex\hbox{\prosima}}

\newcommand{\LCDM}{$\Lambda$CDM\xspace}

\def\sparta{\textsc{Sparta}\xspace}

\def\colossus{\textsc{Colossus}\xspace}

\def\erebos{Erebos\xspace}

\def\vr{v_{\rm r}}
\def\vt{v_{\rm t}}

\def\rtom{R_{\rm 200m}}

\def\rtoc{R_{\rm 200c}}

\usepackage{etoolbox}
\makeatletter
\patchcmd{\NAT@citex}
  {\@citea\NAT@hyper@{\NAT@nmfmt{\NAT@nm}\NAT@date}}
  {\@citea\NAT@nmfmt{\NAT@nm}\NAT@hyper@{\NAT@date}}
  {}
  {}
\patchcmd{\NAT@citex}
  {\@citea\NAT@hyper@{%
     \NAT@nmfmt{\NAT@nm}%
     \hyper@natlinkbreak{\NAT@aysep\NAT@spacechar}{\@citeb\@extra@b@citeb}%
     \NAT@date}}
  {\@citea\NAT@nmfmt{\NAT@nm}%
   \NAT@aysep\NAT@spacechar%
   \NAT@hyper@{\NAT@date}}
  {}
  {}
\patchcmd{\NAT@citex}
  {\@citea\NAT@hyper@{%
     \NAT@nmfmt{\NAT@nm}%
     \hyper@natlinkbreak{\NAT@spacechar\NAT@@open\if*#1*\else#1\NAT@spacechar\fi}%
       {\@citeb\@extra@b@citeb}%
     \NAT@date}}
  {\@citea\NAT@nmfmt{\NAT@nm}%
   \NAT@spacechar\NAT@@open\if*#1*\else#1\NAT@spacechar\fi%
   \NAT@hyper@{\NAT@date}}
  {}
  {}
\makeatother

\newcommand{\rR}{\ensuremath{r/R_{\rm 200m}\xspace}}
\newcommand{\vrV}{\ensuremath{v_{\rm r}/V_{\rm 200m}\xspace}}
\newcommand{\vtV}{\ensuremath{v_{\rm t}/V_{\rm 200m}\xspace}}
\newcommand{\lnv}{\ensuremath{\ln(v^2/v^2_{\rm 200m})\xspace}}

\newcommand{\oasis}{\textsc{Oasis}\xspace}
\newcommand{\fast}{\textsc{Fast}\xspace}
\newcommand{\optimized}{\textsc{Optimized}\xspace}

\journalinfo{The Astrophysical Journal}

\begin{document}


\title{Distinguishing Orbiting and Infalling Dark Matter Particles with Machine Learning}
\author{Ze'ev Vladimir$^{1}$, Calvin Osinga$^{1}$, Benedikt Diemer$^{1}$, Edgar M. Salazar$^{2}$, Eduardo Rozo$^{2}$}
\affil{
$^1$Department of Astronomy, University of Maryland, College Park, MD 20742, USA; \href{mailto:zvladimi@terpmail.umd.edu}{zvladimi@terpmail.umd.edu}\\
$^2$Department of Physics, University of Arizona, Tucson, AZ 85721, USA
}


\begin{abstract}
Dark matter halos are typically defined as spheres that enclose some overdensity, but these sharp, somewhat arbitrary boundaries introduce non-physical artifacts such as backsplash halos,  pseudo-evolution, and an incomplete accounting of halo mass. A more physically motivated alternative is to define halos as the collection of particles that are physically orbiting within their potential well. However, existing methods to classify particles as orbiting or infalling suffer from trade-offs between accuracy, computational cost, and generalizability across cosmologies. We present an efficient, yet accurate, supervised machine learning approach using decision trees. The classification is based on only the particle radii and velocities at two epochs. Compared to detailed analysis of particle trajectories, we find that our model matches the classification of 97\% of particles. Consequently, we are able to quickly and accurately reproduce the density profiles of the orbiting and infalling components out to many virial radii. We demonstrate that our model generalizes to a significantly different cosmology that lies outside the training dataset. We make publicly available both our final model and the code to train similar models.
\end{abstract}



\section{Introduction}\label{sec:intro}

In a \LCDM Universe, halos are the key building blocks of structure within which galaxies form \citep{rees_77}. Quantifying this general picture of matter in the Universe requires that we define some measure of what belongs to a halo. For example, halo masses allow us to summarize the abundance and growth of structure via mass functions \citep{press_74} and underpin our understanding of the galaxy-halo connection \citep{wechsler_18}.

The conventionally favored solution has been to define halos with spherical overdensity (SO) boundaries \citep{lacey_94}, meaning radii chosen to enclose some specific overdensity, e.g. an overdensity of $200$ times the mean or critical density of the Universe (denoted $\rtom$ and $\rtoc$, respectively). While measuring any definition of halo mass is observationally challenging, SO definitions can at least be easily measured in simulations, namely from spherically averaged density profiles. However, SO definitions suffer from a number of serious drawbacks. First, they do not include the outer regions of all orbits, leading to erroneously classified ``backsplash'' subhalos (and satellites) outside the alleged halo boundary \citep[e.g.,][]{balogh_00, mamon_04, knebe_20, diemer_21_subs} which contributes to the phenomenon of assembly bias \citep{gao_05_ab, villareal_17, mansfield_20_ab}.
Second, SO radii and masses partially evolve via pseudo-evolution due to the evolving background density of the Universe \citep{cuesta_08, diemer_13_pe}. And third, a sharp halo boundary leads to an nonphysical break in halo models of the large-scale distribution of matter \citep[e.g.,][]{cooray_02, garcia_21}. 

These issues have motivated a search for alternative, more physical definitions. One option is the splashback radius, which is defined to include the orbits of the most recently accreted particles and subhalos, avoiding the issue of backsplash halos and pseudo-evolution \citep{diemer_14, adhikari_14, more_splashback_2015}. However, due to the distribution of orbital properties, splashback radii are not uniquely defined either, and measuring them can be hard for halos with few particles. More fundamentally, the splashback radius still represents a ``hard'' boundary, which leads to some undesirable effects. Besides subtle questions of subhalo assignments \citep[e.g.,][]{garcia_halo_2019}, halos can be far from spherical, even out to large radii \citep[e.g.,][]{jing_02, mansfield_17}. 

While gravitational boundness is often used to define the matter that constitutes a halo, it is ill-defined in a Universe without a uniquely defined potential at infinity \citep{behroozi_13_unbound, voit_25}. One commonly used boundary-free definition is the set of particles within a certain ``linking length'' of one another, an algorithm called friends-of-friends \citep[or FOF,][]{davis_85}. While simple computationally, there is no clear physical motivation for FOF halos, especially since the linking length is an arbitrary parameter and does not correspond to a particular SO definition \citep{more_11_fof}.

Based on particle dynamics, we can devise a more natural, boundary-free definition that is physically motivated even for non-spherical halos: a halo is the collection of particles orbiting within its potential well. A particle's first pericentric passage serves as a physically motivated criterion to distinguish when it leaves the infalling population and begins orbiting the halo. This definition has recently been explored as a guiding principle for understanding halos and their density profiles \citep{diemer_22_prof1, diemer_23_prof2, diemer_25_prof3, garcia_better_2022, salazar_dynamics-based_2024}. While the orbits of dark matter particles are not observable, the satellite population can at least approximately be understood as a mixture of orbiting and infalling populations \citep{adhikari_21, aung_phase_2021}.  However, some pressing questions remain, for example how to define subhalos in the orbiting-infalling framework or whether orbiting-only halos exhibit a more universal mass function. 

Such investigations are necessarily based on a classification of particles as orbiting or infalling, often for billions of particles. A simple cut in $r-v_r$ space (radius and radial velocity, respectively) is insufficient because a negative radial velocity does not provide enough information to decide whether a particle is on its first infall or has already experienced an apocenter. Information such as the time when particles fell into a halo helps but does not completely separate the populations \citep{garcia_better_2022}. Thus, \citet{diemer_22_prof1} presented an algorithm that considers the full historical information of each particle's orbit. While reliable, this algorithm is computationally intensive and cannot easily be applied to the largest simulations, where often only a subset of particles or snapshots can be stored \citep[e.g.,][]{garrison_18}. 

In this work, we introduce a new classification scheme using a machine learning model. Our model requires only a small subset of a particle's dynamical history, achieving computational efficiency comparable to a phase-space cut, while maintaining excellent agreement with trajectory tracking. We find that a modest set of input variables is sufficient, namely the phase-space of $r$, $\vr$, and tangential velocity $\vt$ at two epochs. The machine learning model identifies complex relationships in the particle's parameters and reproduces decisions based on full orbits in nearly all cases. The model also generalizes to cosmologies other than the one it was trained on. Our algorithm is fast and can be applied to arbitrarily large datasets, opening the door for further explorations of halos as collections of orbiting particles.

The paper is organized as follows. In Section \ref{sec:methods}, we describe our simulations, our training dataset, and the training of our model. In Section \ref{sec:results}, we analyze the performance of the model in various regions of the input parameter space as well as with density profiles. We further interpret the physics behind the model's decisions in Section \ref{sec:discussion} using feature importance and compare the model to other approximate methods using cuts in kinetic-energy space. We summarize our findings in Section \ref{sec:conclusion}.

\section{Methods}\label{sec:methods}

\subsection{Simulations}\label{sec:sims}

We use six $N$-body simulations from the \textit{Erebos} suite \citep{diemer_14, diemer_universal_2015}, which model the same $\Lambda$CDM cosmology, with the WMAP7 parameters, that was used in the Bolshoi simulation \citep[$\Omega_{\rm m} = 0.27$, $\Omega_{\rm b} = 0.0469$, $\sigma_8 = 0.82$, $n_{\rm s} = 0.95$,][]{komatsu_seven-year_2011}. The simulation box sizes range from $63 \: {\rm Mpc}$ (L0063) to $2000 \: {\rm Mpc}$ (L2000), each simulated with $1024^3$ particles. Finally, we use the L0125, L0250, and L0500, $1024^3$ particle simulations with a {\it Planck}-like cosmology \citep[$\Omega_{\rm m} = 0.32$, $\Omega_{\rm b} = 0.0491$, $\sigma_8 = 0.834$, and $n_{\rm s} = 0.9624$,][]{planck_collaboration_planck_2014} to assess how well the model generalizes to other cosmologies. For all simulations, we use the phase-space halo finder \textsc{Rockstar} \citep{behroozi_rockstar_2013} to extract halo information from each simulation, with halo trees constructed using the \textsc{Consistent-Trees} code \citep{behroozi_gravitationally_2013}. 

We use the orbiting-infalling classifications based on the algorithm of \citet{diemer_22_prof1}, which is implemented in the \sparta framework \citep{diemer_splashback_2017, diemer_splashback_2020}, as our fiducial definition. \sparta labels a particle as orbiting if it has undergone one pericenter and as infalling otherwise. A particle is deemed to have had a pericenter when its radial velocity switches from negative to positive. To avoid spurious pericenters caused by noise, an additional constraint is imposed: the first passage requires the angular direction, defined as $\phi(t)=\hat{\textbf{r}}(t)\cdot\hat{\textbf{r}}(t_{ini})$, to be less than 0.5 at the snapshot before or after the particle switches in radial velocity. The majority of pericenters are determined via these conditions, although the algorithm also handles some special cases \citep[for details, see][]{diemer_22_prof1}. As a refinement of the original algorithm, we enforce an additional maximum physical velocity of $1.5\ V_{\rm 200m}$ at the particle's pericenter, where $V_{\rm 200m} =\sqrt{G M_{\rm 200m} / R_{\rm 200m}}$ and $M_{\rm 200m}$ is the mass of the halo enclosed within $R_{\rm 200m}$. This condition prevents rare particles with excessively high velocities from being assigned pericenters, as they are unlikely to remain bound to the halo. We do still see orbiting particles with larger velocities in our datasets as this condition is only applied at the particle's first pericenter.

We emphasize that although we treat the labels provided by \sparta as ground truth for the purposes of training our model, they are not strictly ``truth'' because rare ambiguities exist regarding the maximum distance at which a pericenter can occur and because some particles that are true fly-bys can still be assigned a pericenter. These particles tend to be outliers that arise from unusual conditions and should not significantly affect the training nor evaluation of the model.

\begin{figure*}
    \centering
    \includegraphics[width=\textwidth]{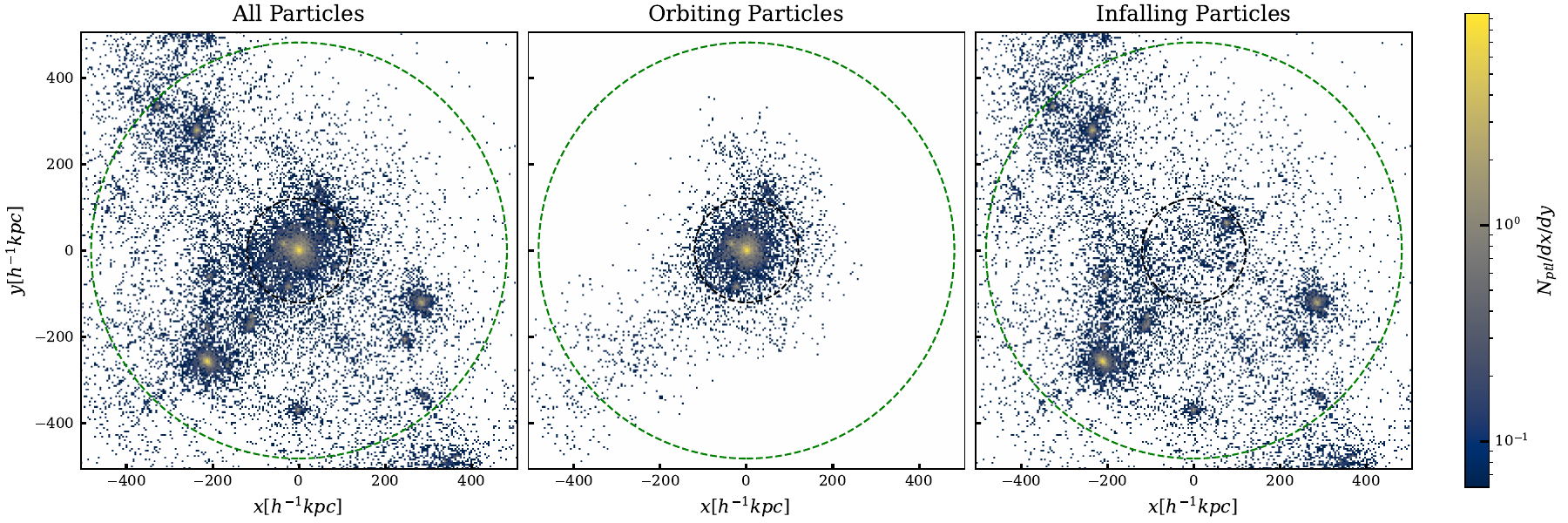}
    \caption{The distribution of particles for a randomly selected dark matter halo from the Bolshoi L0063 simulation (left), separated by \sparta's classifications for orbiting (middle) and infalling (right) particles. The black circle marks $R_{\rm 200m}$ and the green circle is the radius out to which we search for particles at $4 R_{\rm 200m}$. As expected, orbiting particles preferentially exist near the halo center while distant particles tending to be infalling. A non-negligible fraction of orbiting particles, 18.5\% in this case, exist outside of $R_{\rm 200m}$ for this halo, agreeing with other works \citep{diemer_splashback_2017, diemer_splashback_2017-1}.}
    \label{fig:halo_slice}
\end{figure*}

An example distribution of orbiting and infalling particles as determined by \sparta is shown in Fig. \ref{fig:halo_slice}. There are clear populations of orbiting particles that exist outside of $R_{\rm 200m}$, and infalling particles inside $R_{\rm 200m}$. This emphasizes how strict radial boundaries mischaracterize what particles belong to a halo \citep{more_splashback_2015}.

\subsection{Training Dataset}
\label{sec:dset_create}

We consider all particles within $4 R_{\rm 200m}$ of each halo for our dataset, as \citet{diemer_splashback_2017-1} found that this radius contains the vast majority of orbiting particles. We limit our analysis to host halos with a minimum of 200 particles. We exclude subhalos because it is challenging to consistently determine if their constituent particles have a pericenter \citep[e.g.,][]{han_resolving_2012}, which is required by our definition of orbiting.

Simulations with smaller volumes contribute significantly more particles to the overall dataset than larger ones. This phenomenon arises from the higher clustering of halos in small-volume simulations that leads to significant overlap in the particles found within the search radii of the halos, such that individual particles can be counted multiple times. For example, the L0063 simulation has $\sim$11 times as many particles within $4 R_{\rm 200m}$ of all halos as the L2000 simulation. We randomly remove halos from the smaller boxes until each simulation contains an approximately equal number of particles to prevent biasing the model towards these types of halos. We then randomly select 75\% of the halos for the training dataset, reserving the rest for testing. Our final training dataset consists of $1.82 \times 10^9$ particles.

We calculate the particles' physical velocities and positions relative to the center of the host halo to use as model inputs. Due to the average isotropy of the halos we can capture most of the information about a particle's physical velocity in terms of its radial and tangential components. The physical velocity of each particle is $\Vec{v}=\Vec{v}_{\rm ptl}-\Vec{\rm v}_{\rm halo}+\Vec{v}_{\rm hub}$ (where $\Vec{v}_{\rm hub}=H(z)\times r$ is the Hubble velocity), the radial velocity is $v_{\rm r}=\Vec{v}\cdot\hat{r}$, and the tangential velocity is $v_{\rm t}=\sqrt{v^2-v_{\rm r}^2}$. With the halo's $R_{\mathrm{200m}}$ and $V_\mathrm{200m}$ at the current snapshot, we normalize the particles' radii and velocities respectively. The radius, radial velocity, and tangential velocity from an initial snapshot are the first three input parameters for our model.

The remaining inputs are the same three parameters (\rR, \vrV, and \vtV) evaluated at a past snapshot. The addition of the past trajectory information allows the model to distinguish between particles that converged to the same phase-space region at the current snapshot despite traveling distinct paths. To maximize the the range of halos present in the simulations we chose our primary snapshot to be at $z\sim0$. We have tested time separations between the primary and past snapshot of 0.5, 0.75, 1, and 2 dynamical times (where $t_{\rm dyn} = 2R_{\rm 200m} / V_{\rm 200m}$). We find that the model performs best with the snapshots separated by 1.0 dynamical time, with the present snapshot at $z = 0.031$ and the past snapshot at $z = 0.567$. We tested models using no past snapshots which led to noticeable decline in performance and multiple past snapshots which provided negligible improvement, indicating that two snapshots are sufficient to accurately classify orbiting and infalling particles. A side effect of using multiple snapshots is that some particles are not within our search radius both at the present and the past snapshot, or the halo they belong to in the current snapshot did not exist in the past. These particles form a significant population, and we include them in our dataset with an indication that they are ``missing" values for their past parameters. The model still learns from these particles (see Section~\ref{sec:ml_interp}).

\subsection{Model Training}\label{sec:model_train}

For our machine learning model, we select XGBoost \citep{chen_xgboost_2016} because it is easily implemented, scalable to large datasets, and interpretable. XGBoost makes classifications with an ensemble of gradient-boosted decision trees. Each decision tree consists of a series of binary splits based on the particle's input values (e.g., $\rR  \leq 0.5$) until it reaches the final node, which assigns a probability of whether the particle is orbiting or infalling. The final ensemble decision is a combination of each of the tree's probabilities, weighted by the performance of the tree during training. Since each decision tree consists of a series of simple splits, the model's decisions can be easily analyzed in depth, providing insights into the particle populations that constitute the halo.

Our model consists of 100 decision trees, each with a maximum of four decision splits. The number of decision trees constrains overfitting at the ensemble level, while the number of decision splits regulates overfitting within individual trees. Increasing the number of trees or splits generally amplifies overfitting, whereas reducing them mitigates it. However, too few trees or splits can lead to underfitting, limiting the model's ability to capture relevant patterns in the data. We found that 100 decision trees with four splits each provide the model with sufficient complexity to make accurate classifications but retain enough flexibility to generalize to other simulations.

To train the decision trees, XGBoost uses gradient boosting, constructing each tree to minimize the errors from the prior tree. XGBoost provides user settings that control overfitting and training time, called the learning rate and subsample rate. We use the default learning rate of 0.3, which limits the impact of each new tree's prediction on the ensemble classification. Instead of the default subsample rate of 1, we use a rate of 0.5, so each decision tree only receives 50\% of the dataset. This results in shorter training times and reduces the risk of overfitting, as each decision tree is trained on a random subset of data. During training, XGBoost is able to learn from our missing feature values, such as the absence of past particle properties in prior snapshots, incorporating this information into its classifications (Section~\ref{sec:dset_create}). 

During initial training, the model prioritizes classifying infalling particles, as they comprise a larger fraction of the dataset, while largely ignoring orbiting particles outside regions of high concentration. To offset this imbalance, we use the ``scaled position weight parameter'', which allows us to control how much weight XGBoost places on classifying infalling or orbiting particles correctly. We use the ratio of infalling to orbiting particles as our weight to ensure both populations contribute equally to the model. Since our dataset has more infalling particles, the model is incentivized to more heavily weight correct classifications of orbiting particles. We calculate the exact value to be $1.90$. Finally, we use the binary logistic loss function, common in binary classification problems, to measure the model's accuracy during training.

\section{Results}
\label{sec:results}

\begin{figure*}[t!]
    \centering
    \includegraphics[width=\textwidth]{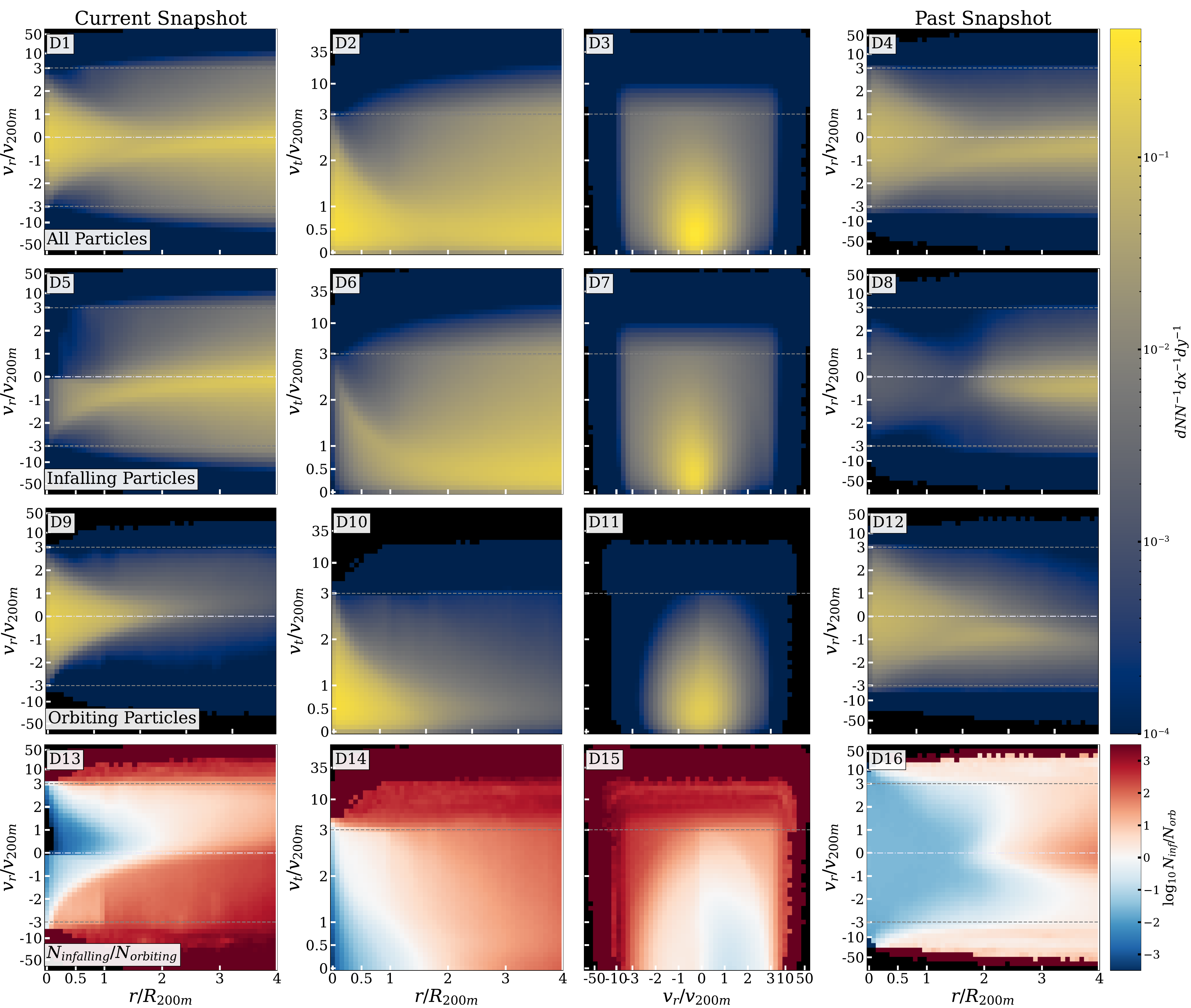}
    \caption{Phase-space distribution of all (top row), infalling (second row), and orbiting particles (third row) from the test dataset, and the ratio between infalling and orbiting (bottom row). The first three rows are normalized by the total number of particles in the dataset and the size of each bin (black signifies an empty bin). The first three columns correspond to the combinations of radii and velocities at the current snapshot, whereas the final column depicts their past radii and radial velocities. We linearly bin the velocity distributions in the inner regions to improve visibility, with logarithmic bins elsewhere (the transition is indicated with gray dashed lines). Each panel is labeled in the top left corner to facilitate discussions in the text. Infalling particles typically reside at large radii and slightly negative radial velocities (D5--D7). Conversely, orbiting particles are concentrated in the halo's center, with larger tangential velocities and an average radial velocity near zero (D9--D11). In the previous snapshot, infalling particles tend to exist at large radii with small radial velocities (D8), as they fail to reach the halo and complete a pericenter before the present snapshot. This is also reflected in the suppressed occupation of orbiting particles in the same region of D12. Regions in phase-space that contain approximately equal numbers of infalling and orbiting particles (white areas in D13--D16) present a challenge for our classifier.}
    \label{fig:ptl_dist}
\end{figure*}

In this section, we evaluate the accuracy of our model's classifications by comparing them against classifications made by the \sparta-based particle trajectory method in the validation halo sample (Section~\ref{sec:dset_create}). We compare the predictions as a function of phase-space parameters (Section~\ref{sec:phase_space}) and in stacked density profiles (Section~\ref{sec:dens_profile}). 

\subsection{Accuracy Across Phase Space}
\label{sec:phase_space}

\begin{figure*}
    \centering
    \includegraphics[width=\textwidth]{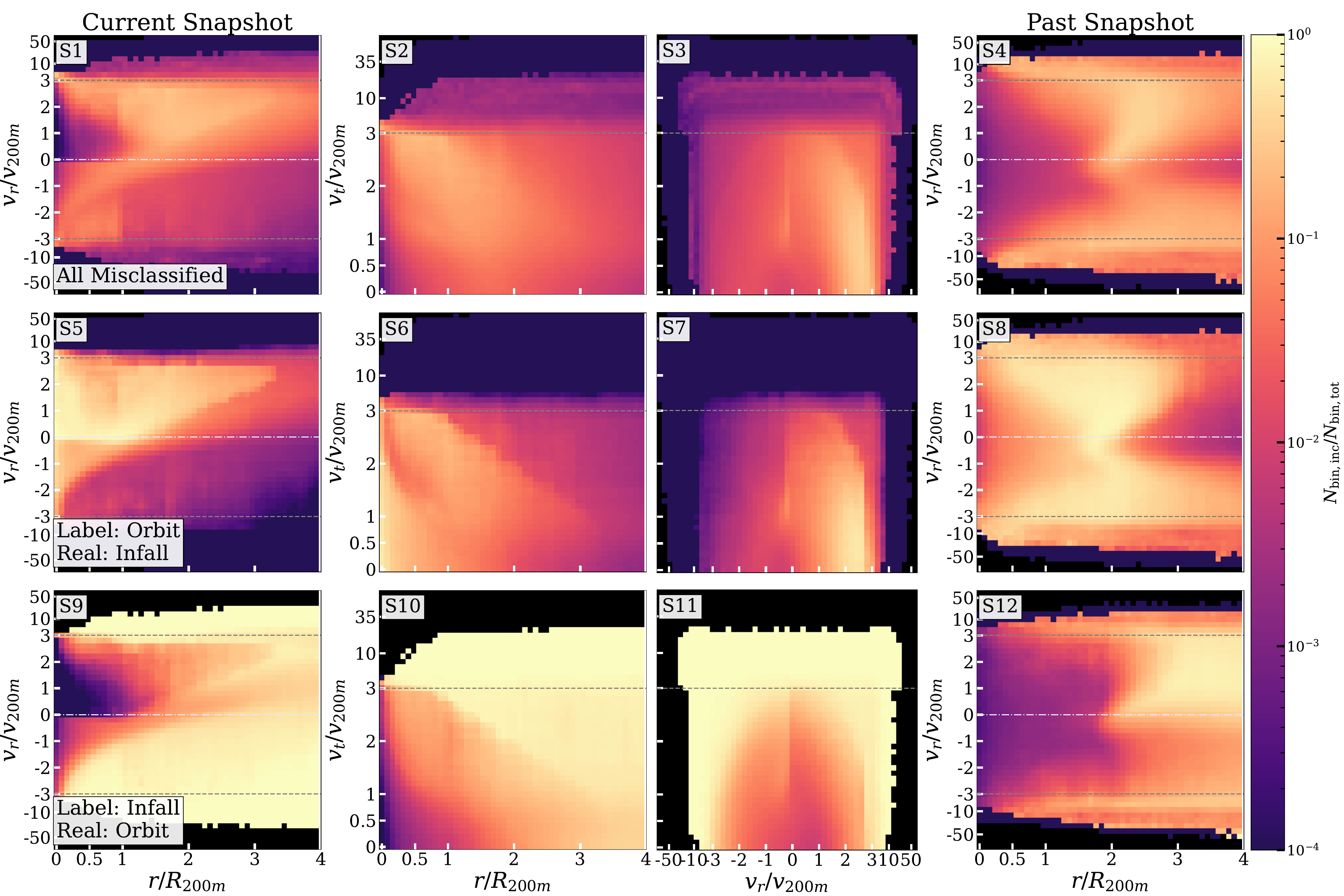}
    \caption{Same as Fig. \ref{fig:ptl_dist} but showing the fraction of misclassified particles in each bin. Areas that have a high concentration of one population correspondingly have low misclassification rates. We suspect that interactions with subhalos lead to the population of infalling particles with a high misclassification rate beyond $0.5R_{\rm 200m}$ and with positive radial velocities (S5). Orbiting particles located outside their areas of high concentration are often misclassified since they are a small, outlier population (S9--S11).}
    \label{fig:miss_class}
\end{figure*}

\begin{figure*}
    \centering
    \includegraphics[width=\textwidth]{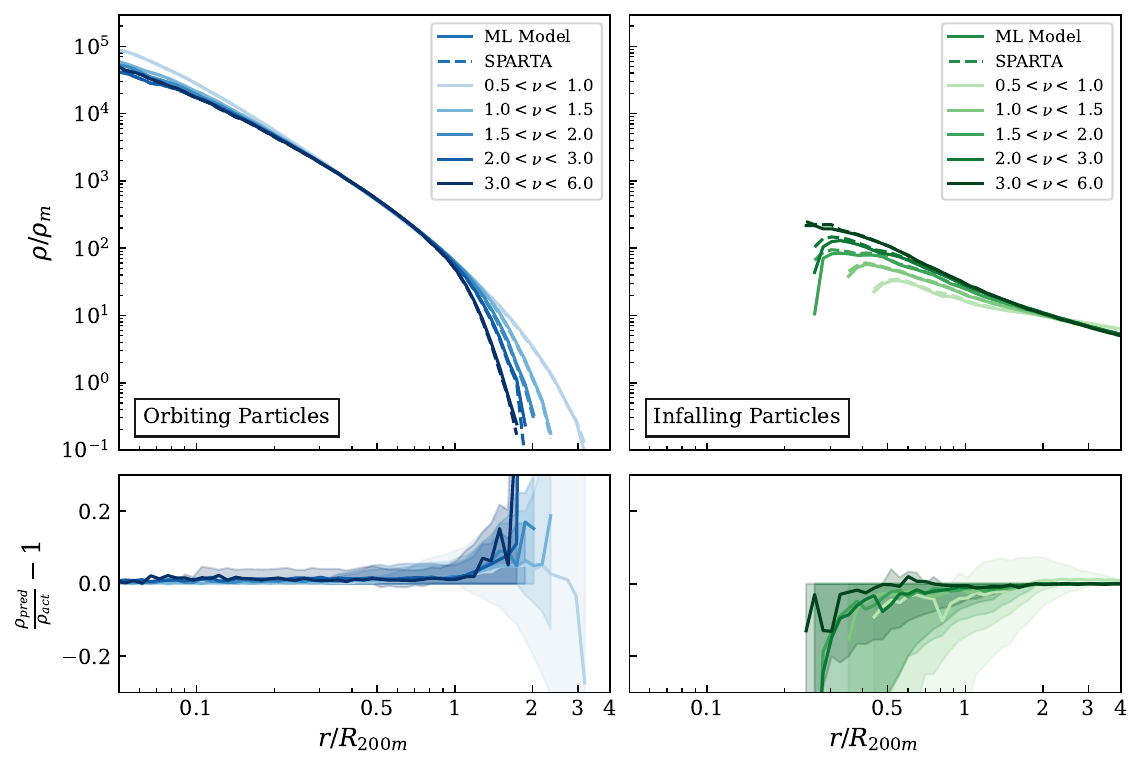}
    \caption{Median orbiting (left), and infalling (right) density profiles of all particles from the WMAP7 test dataset. We bin halos by peak height and hide radial bins where fewer than 50\% of the total number of halos contain particles, with darker colors representing larger peak heights. The solid lines are density profiles from the model's classifications, and the dashed lines are the profiles from \sparta. The bottom panels show the corresponding ratio between the median profiles, with the shaded region representing the 16th and 84th percentile of the ratios between the individual halo profiles. Since both our total profiles and SPARTA's profiles are constructed from the raw particle count in each radial bin, they must agree, as confirmed in the left panel. The model accurately picks up on the key characteristics of the orbiting and infalling profiles across masses. The increased errors for larger peak heights are largely results of a smaller population of halos.}
    \label{fig:dens_prf}
\end{figure*}

In Fig.~\ref{fig:ptl_dist}, we show the phase-space distribution of infalling and orbiting particles as well as their ratio. We quantify the model's classification accuracy across phase space in Fig.~\ref{fig:miss_class}. In this section, we discuss the high-density regions in Fig.~\ref{fig:ptl_dist} that strongly influence the model, as well as the regions with the highest and lowest misclassification rates in Fig.~\ref{fig:miss_class}. We have labeled each panel to facilitate the discussion.

Infalling particles primarily reside at large radii with slightly negative radial velocities (D5) and low tangential velocities (D6). However, their distribution exhibits significant scatter with very large tangential velocities stretching beyond $10 \vtV$ (D7). In contrast, orbiting particles are concentrated at the center of the halo (D9), with higher tangential velocities at smaller radii (D10) and a compact, approximately-Gaussian radial velocity distribution centered around zero but slightly biased toward positive velocities (D11). Both populations overlap on the outskirts of the halo ($0.5$--$1.5 R_{\rm 200m}$) as indicated by the ratio between the populations being approximately unity (D13--D16), demonstrating the challenge in separating them with a single radial threshold. The population ratio exhibits a sharp, seemingly nonphysical transition at $R_{\rm 200m}$ (D13), shifting from a balance of orbiting and infalling particles to a primarily infalling population. This arises because \sparta begins tracking a small population of particles only after they were already inside the halo and so it assigns them an orbiting classification even if a pericenter had not been directly observed. These particles make up a small fraction of the orbiting population and appear strongly here due to the low overall concentration of particles in this region.

There are outlier orbiting particles with extreme velocities reaching $\pm 10 \vrV$ and $10 \vtV$ (D9--D12). A large portion of the orbiting particles at large radii with positive radial velocities are ``fly-bys'', i.e., particles that have had a pericenter but do not remain bound to the halo. The remaining particles are truly orbiting and eventually return to the halo. These outlier regions contain few particles and thus do not significantly impact the training of the model nor our evaluation of its performance.

At the past snapshot (right column of Fig.~\ref{fig:ptl_dist}), the orbiting and infalling particles are distributed similarly to the present snapshot, albeit with three key differences. First, not all particles have a second snapshot (Section~\ref{sec:dset_create}), reducing the overall phase-space density. Second, particles in the past snapshot have a broader range of radial velocities at smaller radii. This phenomenon reflects the broad range of paths that particles can take to arrive at the present snapshot's configuration. Third, distant infalling particles with smaller radial velocities ($\rR > 1.5$ and $-0.5<\vrV<0.25$ in panel D8) typically cannot reach the halo in time to experience a pericenter. This manifests as a low-density region in panel D12, splitting the orbiting particles into two distinct tails at large radii.

The model prioritizes accuracy in regions with high particle concentration, with misclassification rates of $\leq 1\%$ (Fig.~\ref{fig:miss_class}). However, in phase-space regions where the distributions of the two populations overlap, the model's overall accuracy is expected to be lower to balance accuracy across both populations. We focus on two examples of these regions: outlier orbiting particles and infalling particles with relatively large radii and positive radial velocities.

Orbiting particles with rare phase-space parameters are frequently misclassified due to their low abundance (D9). In those regions infalling particles are more prevalent and tend to dominate the orbiting population (D5). Consequently, the misclassification rates of orbiting particles in these regions are outliers that do not significantly affect the model's overall performance. 

The second population consists of infalling particles outside the center of the halo ($0.5<R_{\rm 200m}<1$) with outgoing radial velocities, manifesting as an arc in panel S5. These radii and radial velocities are also characteristic of orbiting particles approaching their apocenters, contributing to increased misclassification in this region. We suspect that subhalo interactions deflect a small fraction of infalling particles into this region. However, the low occupation of infalling particles in this region mitigates the impact of the high misclassification rates on the overall performance.

\subsection{Accuracy of Density Profiles}
\label{sec:dens_profile}

One of the primary goals of the separation of particles into orbiting and infalling was to understand the physical nature of the respective density profiles \citep{diemer_22_prof1}, making density profiles a crucial benchmark for our particle classification method. In this section, we compare the model's density profiles with \sparta's for the WMAP7 simulations on which the model was trained. We then evaluate the model's ability to generalize to other cosmologies.  

We compare the density profiles from our model's classifications with the profiles from \sparta in Fig.~\ref{fig:dens_prf}. To identify potential mass biases in the model's predictions, we further split the profiles by peak height, $\nu_\Delta$, which reflects the statistical significance of halos in the linear overdensity field. We define $\nu_\Delta=\delta_c/\sigma_{M}(M,z)$, where $\delta_c=1.686$ is the overdensity threshold used by the spherical top-hat collapse model \citep[]{gunn_72} and $\sigma_{M}(M,z)$ is the variance of the linear power spectrum measured within spheres of the Lagrangian radius of the halo. The orbiting component dominates the overall profile at small radii, gradually decreasing until a sharp cutoff. On the other hand, the infalling profile's contribution is negligible near the center and overtakes the orbiting contribution at large radii. For a detailed analysis of the orbiting and infalling density profiles, refer to \citet{diemer_22_prof1} and \citet{salazar_dynamics-based_2024}.

The profiles from our model and \sparta closely agree (see difference panels in Fig.~\ref{fig:dens_prf}). Our model reproduces the orbiting density profiles from \sparta within 5\% out to $R_{\rm 200m}$. Beyond this radius, the sharp decline in the number of orbiting particles causes a rapid increase in scatter, but the shapes of the median profiles are captured well. The infalling profiles exhibit the opposite trend, showing high scatter and lower accuracy within $R_{\rm 200m}$ due to the small particle population, while the ratio between the profiles converges at larger radii. We observe a slight mass dependence in the model's accuracy, with halos of larger peak height showing the greatest discrepancy. We attribute this trend to the smaller number of large halos in the training set.

\begin{figure*}
    \centering
    \includegraphics[width=\textwidth]{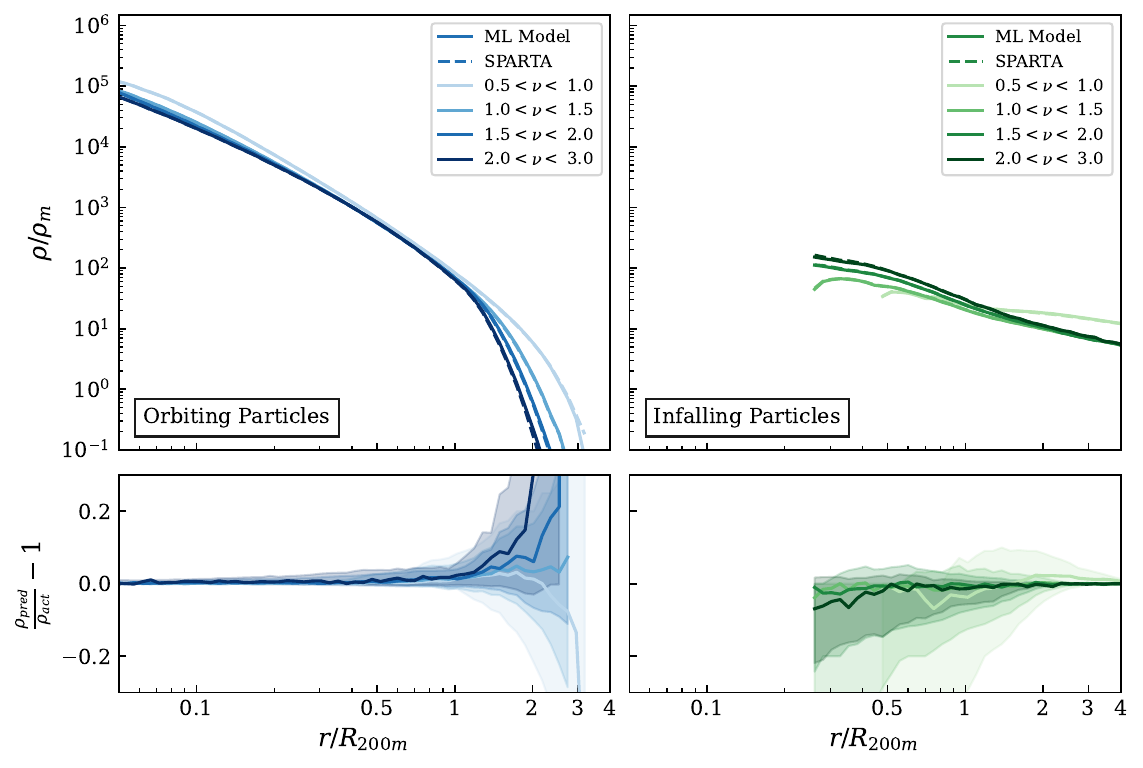}
    \caption{Same as Fig. \ref{fig:dens_prf}, but for the {\it Planck}-like cosmology. Similarly to the fiducial cosmology, the model closely matches the density profiles based on \sparta except for regions with very few particles. Unlike the Bolshoi cosmology simulations, the {\it Planck} cosmology simulations only reach a box size of $500 \: \rm{Mpc}$, which results in too few halos in the $3.0 < \nu < 6.0$ bin.}
    \label{fig:cpla_dens_prf}
\end{figure*}

Despite removing the first-order dependence of our model on halo mass via the scaling of our input parameters, we cannot exclude the possibility that the cosmology has residual effects \citep[e.g., due to slightly different splashback radii;][]{diemer_splashback_2017-1}. To assess if the chosen cosmology does impact our results, we compare the stacked density profile for the {\it Planck} cosmology simulations to \sparta's corresponding profiles in Fig.~\ref{fig:cpla_dens_prf}. We observe similar trends as in Fig.~\ref{fig:dens_prf} in both the orbiting and infalling profile accuracy. The misclassification rates of particles in the {\it Planck} cosmology also matches closely with that of the WMAP7 cosmology. Regions of high and low accuracy are consistent between the two, indicating that the populations within the halos are very similar across cosmology.  

\section{Discussion}
\label{sec:discussion}

\subsection{Interpreting the ML Model}
\label{sec:ml_interp}

Understanding why our model makes its classification choices allows us to assess whether it captures meaningful physical relationships between the particles and their features or relies on spurious correlations. Feature importance is a typical way to interpret a model's choices, providing useful insights into how individual features contribute to a model's predictions. We caution that feature importance cannot establish causal relationships between the model's predictions and physical processes. However, they are still useful to understand correlations and for identifying populations, in our case of particles, of interest to examine more thoroughly.


We use SHapley Additive exPlanations (SHAP) values \citep{lundberg_local_2020} to probe how the model classifies various populations of interest. SHAP values utilize a game theory approach to understanding how the input variables of each particle influence the model's classification of that particle. They explain the contribution of each feature by quantifying the change in the output classification when each feature is included or excluded. Feature values that the model considers critical for classification have large SHAP values. We define the SHAP values such that positive values indicate parameters the model associates with orbiting particles, while negative values correspond to infalling ones.

\begin{figure*}
    \centering
    \includegraphics[width=\textwidth,height=10cm]{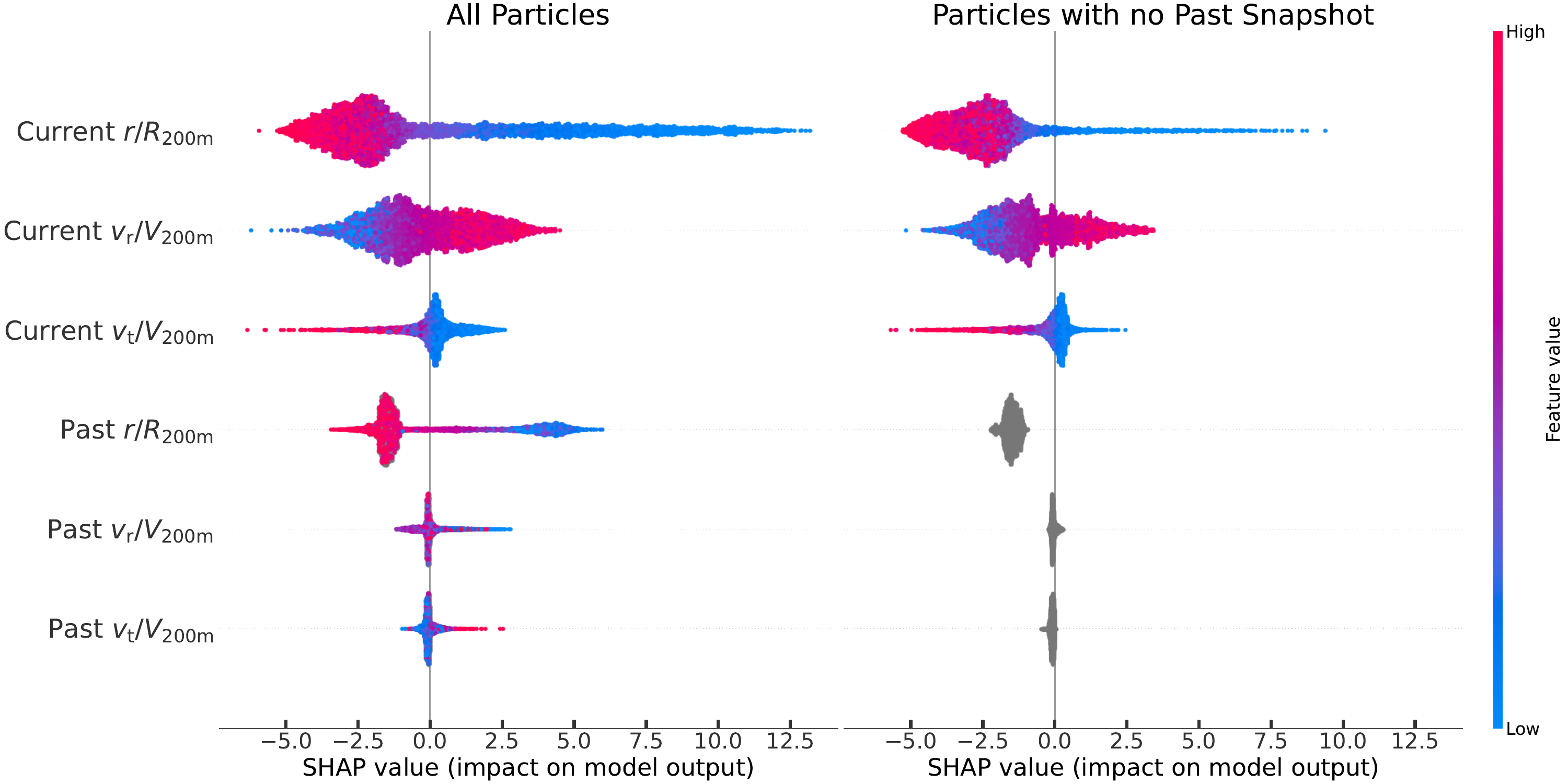}
    \caption{Feature importance for each particle in the model's classifications, as captured by SHAP values. We take two random samples of 5,000 particles from our testing dataset, one from all particles and one from particles that do not have past snapshot information. We examine how the model weighted each feature's value in its classification. Negative SHAP values mean that a feature is indicative of infalling particles, with positive values indicating orbiting. The color scale represents the range of values for each feature rather than exact numbers, with red points indicating higher values and blue points indicating lower values for a specific particle's feature. Grey points indicate that the values for that feature are not present, which happens for particles that were not within the search radius in the past snapshot. The radius of the particle at the current and past snapshot has the largest influence on the model's classification for both populations, indicated by the larger SHAP values the model assigns this feature. The present radial and tangential velocities have similar shapes for both groups while past velocities only significantly contribute with extreme velocities.}
    \label{fig:shap_plt}
\end{figure*}

\begin{figure*}
    \centering
    \includegraphics[width=\textwidth]{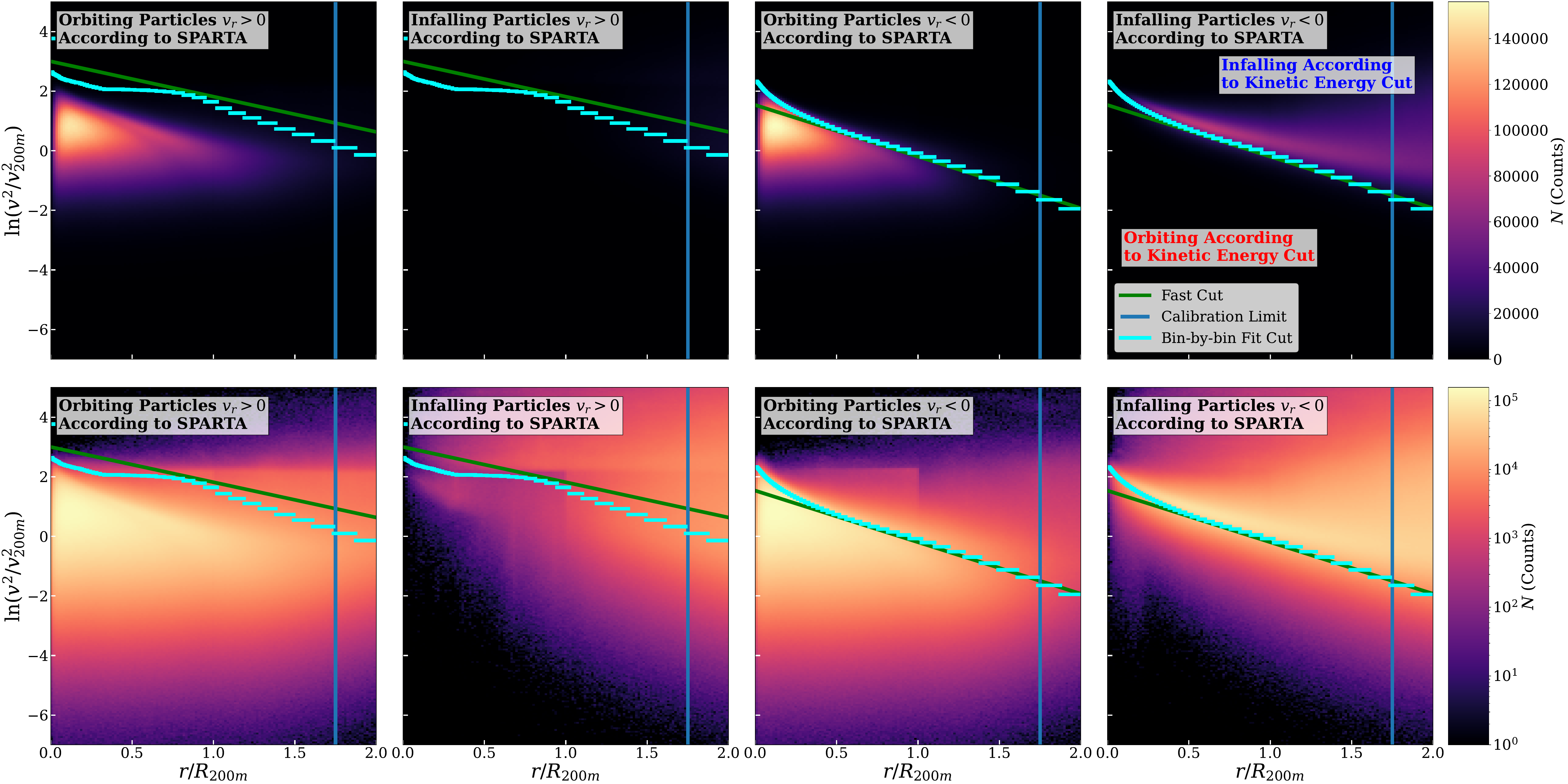}
    \caption{Comparison of kinetic-energy cuts to \sparta's classification. We investigate cuts in linear (top-row) and log (bottom-row) scaled phase-space of $\lnv$ versus $r/R_{200m}$ from all the WMAP7 simulations. The particles are split by their classification according to \sparta and then further by their sign of their radial velocity. The two kinetic-energy cut methods, described in Section~\ref{sec:comparison}, are overplotted as green lines (\fast) and cyan bins (\optimized). The kinetic-energy cuts classify particles below their lines as orbiting and those above as infalling. The blue line indicates the furthest radial extent that particles are considered for the simple kinetic-energy cuts' fit. The bin-by-bin fit of the \optimized method disagrees with the best-fit lines significantly, and naturally leads to a more accurate match to \sparta's classification.}
    \label{fig:ke_cut_dist}
\end{figure*}

A SHAP beeswarm plot (Fig.~\ref{fig:shap_plt}) illustrates the distribution of each feature's importance for each particle along the horizontal axis. An important feature for determining infalling particles will have a higher density on the left, and vice versa for orbiting particles. The color of each point indicates feature values, with red representing larger values, blue indicating smaller ones, and gray signifying missing data. We find that the model places the highest importance on the radius of a particle at both the present and past snapshot, followed by the current radial velocity and the current tangential velocity (Fig.~\ref{fig:shap_plt}). Interestingly, the model only relies on the past velocities when they exhibit more extreme values, indicated by the centering of these parameters around zero. We can attribute it to these parameters providing redundant information unless they are abnormal.

We examine the SHAP values of the population of particles that do not have secondary snapshot parameters in the right panel of Fig.~\ref{fig:shap_plt}. These are particles within the search radius of a halo at the present snapshot but were either outside the search radius or belonged to a halo that did not exist in the past snapshot. These particles lack values for their past snapshot's parameters, which the XGBoost model can interpret and use in its classifications. This population is significant, comprising approximately $45.8\%$ of all particles in our dataset of which $88.8\%$ are infalling, as classified by \sparta. 

Although this population has a similar distribution of SHAP values in the present features, there is a much tighter clustering around zero for the past $\vrV$ and $\vtV$, indicating that they have a smaller impact on the particle's classification. However, unlike in the all-particle sample, the past radius feature is used to push every particle towards an infalling classification. This is physically motivated, given that a particle that was outside the search radius in the past snapshot is unlikely to have experienced a pericenter. The particles in this population that managed to reach the halo's center at the primary snapshot, represented by the blue dots for the current $\rR$ feature, likely consist of two subpopulations. The first are particles that had a large enough velocity to achieve a pericenter between the snapshots and the others would have belonged to halos forming between the past and current snapshots.

Overall, the model seems to classify particles without past information as infalling, as one would expect for particles that were outside the search radius at that time. The only exceptions are particles that belong to halos that did not exist 1.0 dynamical time ago or had fast enough infalling velocities to have a pericenter despite the distance. Only a small subset of this population eventually orbits the halo, reinforcing our decision to limit the snapshot separation to 1.0 dynamical time. Increasing the time separation would increase the number of particles in this category, reduce the relevance of past snapshots, and provide less information on a particle's orbiting classification.

\subsection{Comparison to Kinetic Energy Cut}
\label{sec:comparison}

While our ML algorithm accurately reproduces \sparta's particle classification, one might wonder whether this success could also be reproduced by a simpler criterion such as a kinetic-energy cut. In a soon-to-be-published paper, \citep{salazar_manuscript_2025} argue that a simple kinetic energy cut can be used to quickly separate orbiting from infalling particles.  This type of cut is motivated by the expectation that orbiting particles should be bound and therefore have lower kinetic energies than the infalling particle population.  Here, we wish to determine whether the proposed kinetic energy cuts are also a reasonable match to orbiting/infalling decomposition based on Sparta.  We consider two such algorithms: \fast, which refers to the algorithm of \citet{salazar_manuscript_2025}, and \optimized, which adopts radius-dependent kinetic energy cuts that maximize the agreement with \sparta.

The \fast method forms the basis of the \oasis code, which will be presented in \citet{salazar_manuscript_2025}. Briefly, the distribution of kinetic energies as a function of radius is split in two, according to whether they have positive or negative radial velocities.  As expected, the distribution of particles with negative radial velocities is more or less bimodal, with infalling particles having higher kinetic energies.  \oasis self-calibrates a linear cut in this space to isolate the two components, while further imposing a conservative cut on the kinetic energy of \it positive \rm radial velocity particles to ensure that unusually energetic particles be labeled as non-orbiting (and therefore ``infalling'', even though this is a misnomer for these particles). Finally, particles with $R\geq 2R_{\rm 200m}$ are all classified as infalling.  A more detailed description of the \fast algorithm will be presented in \citet{salazar_manuscript_2025}.

\begin{figure}
    \centering\includegraphics[width=0.5\textwidth]{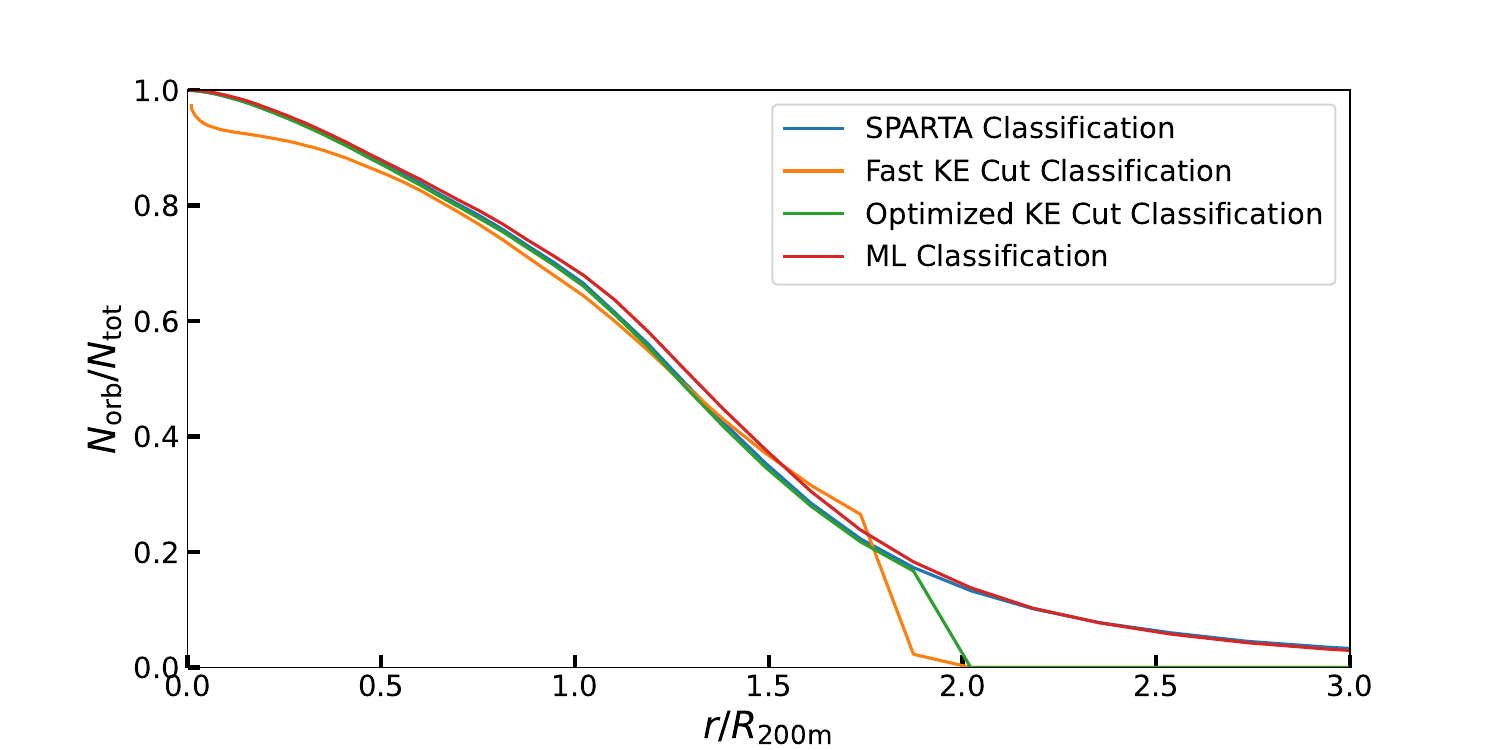}
    \caption{The fraction of orbiting particles at each radial bin according to \sparta (blue), \fast (orange), \optimized (green), and our machine learning model (red). We see that our model matches very closely with \sparta as expected. The kinetic-energy cuts match the general shape of \sparta with \optimized having a closer match due to being directly fitted to \sparta's classifications, but both drop to 0 at $2R_{\rm 200m}$ where we have our hard cutoff for orbiting particles.}
    \label{fig:forb_by_rad}
\end{figure}

Figure ~\ref{fig:ke_cut_dist} compares the \fast method (green lines) to the distribution from \sparta in isolated halos, defined as halos whose most massive neighbor within $2R_{\rm 200m}$ is at most 20\% of the mass of the central halo.  We split the particle population into four samples with positive/negative radial velocities and orbiting/infalling classifications from \sparta. The green lines in Fig.~\ref{fig:ke_cut_dist} show that the linear kinetic-energy cut separates the bulk of the orbiting and infalling populations as classified by \sparta. The figure shows both linear (top) and logarithmic (bottom) scales, highlighting the bulk of the population and outliers, respectively. Figure~\ref{fig:forb_by_rad} additionally shows the fraction of orbiting particles as a function of radius as determined by both algorithms. The \fast and \sparta algorithms are in reasonably good agreement, except for the very inner regions and beyond $2 R_{\rm 200m}$, where \fast classifies all particles as infalling.

\begin{figure*}
    \centering
    \includegraphics[trim={2cm 0 2cm 0},width=\textwidth]{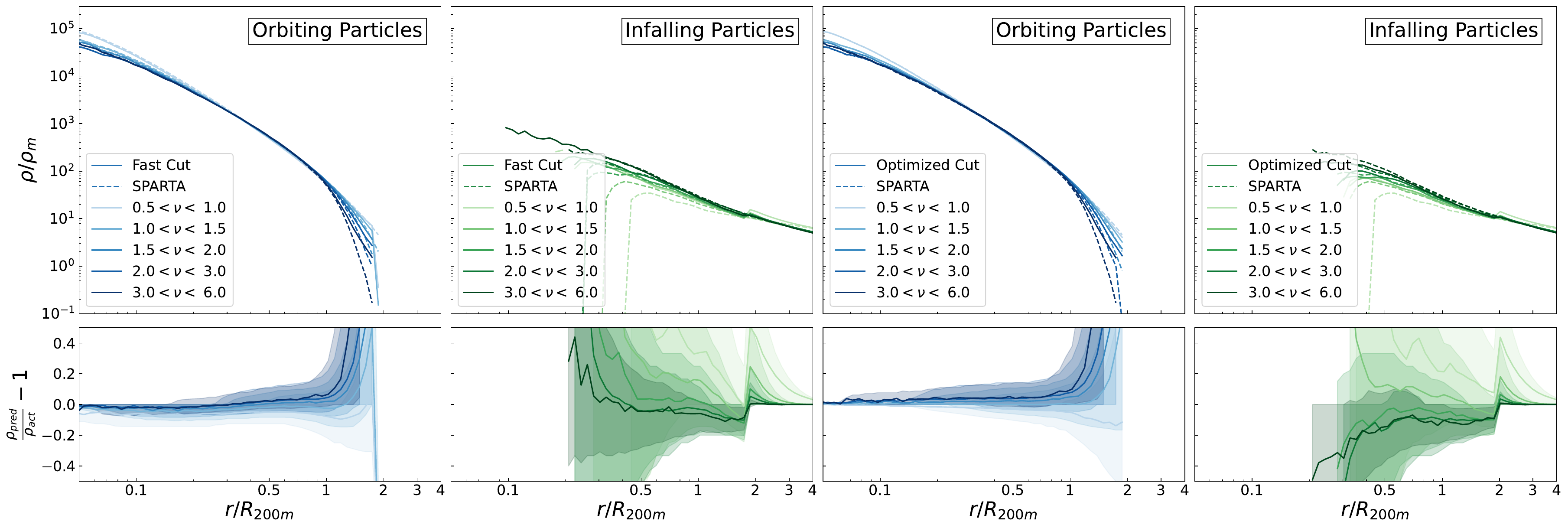}
    \caption{Same as Fig. \ref{fig:dens_prf}, but with the predictions made with the \fast model (left two columns) and with our \optimized model (right two columns). The kinetic-energy cuts used to make the classifications are shown in Fig.~\ref{fig:ke_cut_dist}. Both methods do reasonably well in capturing the concentrations of orbiting and infalling particles for methods that use only one snapshot. However, even within $R_{\rm 200m}$ the errors for all but the smallest halos are much larger than our machine learning model and for the largest halos reach more than 10\%. The kinetic-energy cuts fail to capture the shape of the density profiles beyond $R_{\rm 200m}$ indicated by the different cutoff points present for each mass bin. In addition, the errors in the infalling profiles approach zero at larger radii than for our machine learning model.}
    \label{fig:ke_dens_prf}
\end{figure*}

By contrast, the logarithmic scale for the bottom row of Fig.~\ref{fig:ke_cut_dist} highlights the tails of the distributions, making it easier to see where the \sparta and \fast classifications disagree.  The most important difference is due to a population of particles classified as infalling by \sparta at large radii and with low kinetic energy, i.e the particles that fall below the \fast line in the right-most plot of Fig.~\ref{fig:ke_cut_dist}.  These particles ``boost'' the outskirts of the orbiting profile of halos in the \fast algorithm relative to the profile obtained with \sparta.  This is illustrated in Fig.~\ref{fig:ke_dens_prf}, where we compare the orbiting and infall profiles of halos as determined using \fast and \sparta. The truncation of the \sparta orbiting profiles is much more sensitive to peak height than the truncation observed using the \fast algorithm.  This type of sensitivity is expected: the outer profile is known to be sensitive to accretion rate \citep{diemer_14}, which increases with peak height, suggesting the outer profiles in \sparta are better physically motivated than those obtained using the \fast algorithm. 

However, it is not immediately clear whether this is a limitation of the \fast algorithm in particular or kinetic-energy cuts in general. We have thus also tested the \optimized method,  which corresponds to finding a radius-dependent kinetic energy cut that minimizes the differences with \sparta's classification. Figure ~\ref{fig:ke_cut_dist} shows that the resulting cuts (in cyan) cannot be adequately described as a line. Naturally, the \optimized classification better matches \sparta, but the right columns of Fig.~\ref{fig:ke_dens_prf} demonstrate that differences near the truncation of the orbiting profiles remain. These differences are fundamentally due to the fact that distant, low-energy infalling particles have phase-space positions indistinguishable from orbiting particles near their apocenter, leading phase-space based classifiers to label these particles as orbiting.  However, when adding information from a snapshot that is 1.0 dynamical time earlier, it becomes trivial to separate these particles: infalling particles used to be much further away from the halo, while orbiting particles in the halo outskirts used to be near the halo center at that time.  We have explicitly verified this is the case, explaining why our machine-learning method can adequately distinguish between orbiting and infalling particles in the halo outskirts.

We have also tested whether one can apply the kinetic energy cuts calibrated in the WMAP7 box to our {\it Planck} cosmology boxes.  We find that unlike our machine learning algorithm, the \fast cuts calibrated in WMAP7 do not generalize to the other cosmology.  Indeed, in \citet{salazar_manuscript_2025} we show that kinetic energy cuts scale linearly with matter density (but not other parameters). Interestingly, our single snapshot machine learning method exhibits a similar degradation in performance to that of the \fast method when moving from the WMAP7 simulations to {\it Planck}.  This suggests that the cosmology dependence in the \fast algorithm is caused by the ambiguity in the classification of distant low-energy particles in single-snapshot data highlighted above.  Since these particles are trivially classified upon the inclusion of a second snapshot, our two-snapshot algorithm can be successfully applied in the {\it Planck} simulations even when calibrated using the WMAP7 simulations.

\section{Conclusions}
\label{sec:conclusion}

We have trained the first machine learning model to predict whether particles are orbiting a halo or infalling. The prediction is based on only the particle radii, radial velocities, and tangential velocities at two snapshots. We evaluate the resulting classifications against a more sophisticated algorithm that considers entire orbital trajectories. Our main conclusions are as follows: 

\begin{enumerate}
    \item It is possible to accurately classify particles as orbiting or infalling based on the six input variables, with 97\% accuracy overall.
    The model accurately reproduces the stacked orbital density profiles across halo masses to within 5\% out to $R_{\rm 200m}$.
    \item We demonstrate the model's generalizability by showing that it reproduces stacked density profiles from an out-of-sample simulation with a different cosmology as well as for the training cosmology. This feature relies on multiple snapshot data, meaning that single-snapshot classifiers are necessarily cosmology-dependent.
    \item We interpret the model via feature importance. The most important parameters for classification are the particle's radii at both snapshots, followed by the radial velocities.
    \item We have verified that simple kinetic energy cuts can accurately recover the fraction of orbiting particles as a function of radius in halos. However, such simple classifiers fail to reproduce the full accretion-rate dependence of the orbiting profile outskirts.
\end{enumerate}

Our machine learning model provides accurate, quick, and generalizable classifications for particles in dark matter only simulations. Our tool lays the foundation for possible applications of the orbiting-infalling split as a more general way to characterize halos. However, there are many avenues of research left. For example, we did not explore the possible redshift dependence of our model. We also plan to apply our ML model to large cosmological simulations of galaxy formation.

\section*{Acknowledgments}
This research was supported in part by the National Science Foundation under Grant numbers 2206690. BD is grateful to the Sloan Foundation for their financial support. E.S. and E.R. were supported by NSF grant 2206688.  E.R. is also funded by DOE grant DE-SC0009913. The {\it Erebos} simulations were run on the \textsc{Midway} computing cluster provided by the University of Chicago Research Computing Center, and our analysis was performed on the \textsc{Zaratan} cluster at the University of Maryland. This research extensively used the python packages \textsc{Numpy} \citep{code_numpy2}, \textsc{Scipy} \citep{code_scipy}, \textsc{Matplotlib} \citep{code_matplotlib}, \textsc{Pygadgetreader} \citep{thompson_pygadgetreader_2014}, \textsc{XGBoost} \citep{chen_xgboost_2016}, \textsc{SHAP} \citep{lundberg_explainable_2019}, and \colossus \citep{diemer_18_colossus}.


\section*{Data Availability}

Our code is publicly available at \href{https://github.com/ZevVladimir/ATHENA}{github.com/ZevVladimir/ATHENA} or at \dataset[10.5281/zenodo.17228903]{\doi{10.5281/zenodo.17228903}}, including a description of how to create the datasets needed and how to train additional models. We also provide the json of the XGBoost model used and configuration parameters. The full particle data for the \erebos $N$-body simulations are too large to be permanently hosted online, but they are available upon request. 


\bibliographystyle{mnras}
\bibliography{bib/references, bib/bib_general, bib/bib_structure, bib/bib_galaxies, bib/bib_mine}


\clearpage


\label{lastpage}
\end{document}